\documentclass[amsmath, amssymb, english, aps, prl, showpacs, twocolumn, reprint, floatfix, superscriptaddress]{revtex4-1}
\usepackage[english]{babel}
\usepackage{graphicx}
\usepackage[usenames,dvipsnames]{color}
\usepackage[normalem]{ulem}

\begin{document}

\author{Wei Wu}
\affiliation{Centre de Physique Th\'eorique, \'Ecole Polytechnique, CNRS, Universit\'e Paris-Saclay, 91128 Palaiseau, France}
\affiliation{Coll\`ege de France, 11 place Marcelin Berthelot, 75005 Paris, France}

\author{Michel Ferrero}
\affiliation{Centre de Physique Th\'eorique, \'Ecole Polytechnique, CNRS, Universit\'e Paris-Saclay, 91128 Palaiseau, France}
\affiliation{Coll\`ege de France, 11 place Marcelin Berthelot, 75005 Paris, France}
\author{Antoine Georges}
\affiliation{Coll\`ege de France, 11 place Marcelin Berthelot, 75005 Paris, France}
\affiliation{Centre de Physique Th\'eorique, \'Ecole Polytechnique, CNRS, Universit\'e Paris-Saclay, 91128 Palaiseau, France}
\affiliation{DQMP, Universit\'e de Gen\`eve, 24 quai Ernest Ansermet, CH-1211 Gen\`eve, Suisse}
\author{Evgeny Kozik}
\affiliation{Physics Department, King's College London, Strand, London WC2R 2LS, UK}

\begin{abstract}
We introduce a method for summing Feynman's perturbation series based on diagrammatic Monte Carlo that significantly improves its convergence properties.  
% We use it 
This allows us to %controllably 
investigate in a controllable manner the pseudogap regime of the Hubbard model 
and to study the nodal/antinodal dichotomy at low doping and intermediate coupling.
%in which 
Marked differences from the weak coupling scenario are manifest, such as a higher degree of incoherence 
at the antinodes than at the `hot spots'. 
%and study the nodal/antinodal dichotomy at low doping and intermediate coupling. 
Our results show that the pseudogap and reduction of quasiparticle coherence at the antinode  
is due to antiferromagnetic spin correlations centered around the commensurate $(\pi,\pi)$ wavevector.  
%and that umklapp scattering is essential to this effect.  
In contrast, the dominant source of scattering at the node is associated with incommensurate momentum transfer. 
%Our results show that the reduction of quasiparticle coherence is due
%to spin scattering, but while the associated momentum transfer is commensurate
%at the antinode it is incommensurate at the node. 
%Moreover, 
Umklapp scattering is found to play a key role in the nodal/antinodal dichotomy. 
\end{abstract}

\pacs{71.10.-w, 71.10.Fd, 71.27.+a, 74.72.-h}
%71.10.-w Theories and models of many-electron systems
%71.10.Fd Lattice fermion models (Hubbard model, etc.)
%71.27.+a Strongly correlated electron systems; heavy fermions
%74.72.-h Cuprate superconductors

\title{Controlling Feynman diagrammatic expansions: physical nature of the pseudo gap in the two-dimensional Hubbard model}

\maketitle

\paragraph{Introduction and context.}

Strongly-correlated many-electron systems are a major theoretical challenge. 
%for theoretical condensed-matter physics. 
%Analytical solutions exist only in particular limits. 
Numerical approaches face difficulties brought by the
exponentially large Hilbert space or the fermionic sign problem. 
Among the many questions still open, an outstanding one is  
%and many open questions remain. A particularly important one is 
the nature of the low doping and intermediate to strong coupling regime of the prototypical
two-dimensional Hubbard model. %To make progress, there is a definite need for
Unbiased methods are needed to establish whether key aspects of cuprate
phenomenology, such as the opening of a pseudogap and the associated
nodal/antinodal (N/AN) dichotomy~\cite{timusk1999pseudogap,badoux2016pseudo}, are
intrinsic features of the model, and to settle the 
much debated physical origin of these phenomena. Recently, cluster-DMFT approaches have allowed
significant progress on these issues~\cite{georges1996dmft, maier2005dca,
senechal2004pseudogap, leblanc2015}. However, cluster methods lack fine 
momentum resolution~\cite{gull2010eightsite}, which is crucial in
view of the strong momentum dependence along the Fermi surface, and in
establishing which fluctuations are responsible for the physics~\cite{gunnarsson2015fluct}, e.g.
distinguishing commensurate and incommensurate fluctuations.

A promising
alternative method is the diagrammatic Monte Carlo (DiagMC)
technique~\cite{prokofev1998diagmc,proceedings,kozik2010diagmc}, based on the
stochastic summation of Feynman diagrammatic series %immediately 
directly in the thermodynamic limit, which in principle enables 
controlled solutions with arbitrary momentum resolution. However, for lattice systems, fundamental problems with
series convergence have so far limited its scope of application to modest
couplings, relatively high temperature and/or low density. In particular, the
skeleton series built on the full (interacting) Green's function $G$ can converge
to a wrong answer~\cite{kozik2015luttinger}, whereas the bare series built on the non-interacting Green's
function $G_0$ does not exhibit misleading convergence but typically diverges
in the strongly-correlated regime.      

In this Letter, we introduce an approach that considerably enlarges the
applicability range of DiagMC. It is based on a parametric modification of the
bare diagrammatic series that improves its convergence properties. This
technique allows us to address the pseudogap regime of the 2D Hubbard model at
small doping and intermediate coupling. The high momentum resolution and direct
access to scattering processes in DiagMC allows us to identify the physical origin of the pseudogap and N/AN dichotomy, 
which are shown to result from antiferromagnetic spin correlations.    
%We demonstrate that the scattering both at the nodal and antinodal regions of the Fermi surface is exclusively driven by spin fluctuations. 
At the node, the transfer momentum $\mathbf{q}$ of relevant modes is found to be incommensurate, connecting Fermi surface points. 
In contrast, at the antinode, scattering with commensurate momentum exchange $\mathbf{q}=(\pi, \pi)$ dominates. 
We find that quasiparticles are more incoherent at the antinode than at the `hot spots' ( where the Fermi surface intersects the antiferromagnetic zone boundary),  thus establishing the strong-coupling 
nature of the regime investigated. We show that the umklapp scattering enhanced at large perturbation orders plays a key 
role in this %coherence suppressing. 
suppression of coherence.  
%We also show that the stronger spectral weight reduction at the antinode %as compared to node 
%is mainly due to the \ag{strong-coupling} umklapp scattering enhanced at large pertubation orders and that the 
%van-Hove singularity plays a vital role, \ag{which is clearly distinguished from the conventional weak-coupling umklapp
%scattering theory}~\cite{lehur2009review,furukawa1998umklapp, honerkamp2001weakumklapp}  .

\paragraph{The method: action optimization and recursive evaluation of diagrams.}

\begin{figure*}[!t]
 \begin{center}
\includegraphics[width=0.98\textwidth, height=0.45\textwidth]{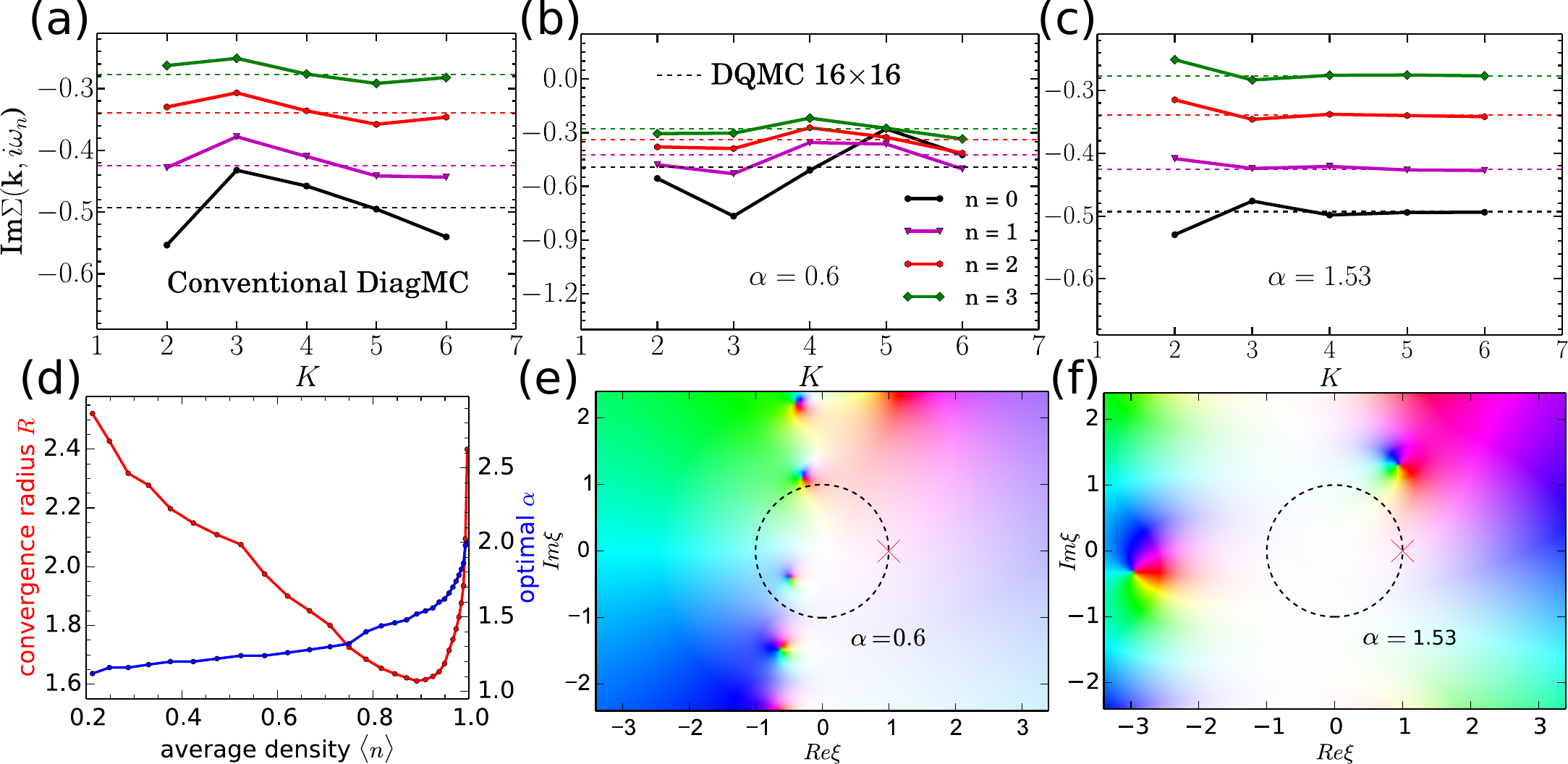}
   \end{center}
   \caption{\label{fig:convergence}
     \emph{Top row}: Imaginary part of the Hubbard model self-energy at the first
     four Matsubara frequencies obtained as a sum of the first $K$ perturbation orders.
     The parameters are $U=4.0$, $t^{\prime}=-0.3$, $\mu=0$, $n\sim0.725$, $T = 0.5$.
     The dashed lines are a benchmark from determinantal QMC simulations on a
     $16\times 16$ lattice (the discrete time interval is $\Delta \tau = 0.0375$
     where the Trotter error is negligible). \textbf{(a)}: Standard series with Hartree diagrams included in bare Green's function, \textbf{(b)}:
     $\alpha$-shifted case with $\alpha = 0.6$, \textbf{(c)}: the optimal case $\alpha
     = 1.53$.  \emph{Bottom row}: Results for the atomic limit $t=0$, $U=4$,
     $T=0.5$, $\mu=0.138$, $n\sim0.725$.  \textbf{(e)} and \textbf{(f)}: Modulus/phase (displayed as
     saturation/hue) map of the self-energy in the complex $\xi$-plane for
     $\alpha=0.6$ and $\alpha=1.53$ respectively. The physical solution is at the
     cross $\xi=1$ on the unit circle.  \textbf{(d)}: Optimal $\alpha$ and
     corresponding maximal convergence radius as a function of the density $\langle
     n \rangle$ in the Hubbard atom.}
   %The convergence of the $\alpha$-optimization on the Hubbard model with
   %$U=4.0$, $t^{\prime}=0.3$, $\mu=0$, $n\sim0.73$, $T = 0.5$ and corresponding
   %pole map in the atomic limit. To illustrate the convergence behaviour, the
   %imaginary part of the self-energy at small frequencies are shown as functions
   %of perturbation order $K$. Determinant QMC simulation on $16\times 16$ lattice
   %is carried out as benchmark. The DQMC simulation is done with discrete time
   %interval $\Delta \tau = 0.0375$, where the Trotter error is negligible.
   %\textbf{(a):} Without $\alpha$ -shift. In obedience to the conventional way of
   %renormalizing the diagMC series, diagrams with bold tadpoles are removed here.
   %\textbf{(b):} $\alpha$ -shifted case with $\alpha = 0$, where bad convergence
   %is seen. See below the corresponding pole map. \textbf{(c):} The optimal case
   %$\alpha = 1.53$, the convergence is greatly improved here compared to the
   %no-shift and $\alpha = 0$ cases. In the corresponding pole map below, we can
   %see the convergence radius is maximized in this case.  \textbf{(d):} Darker
   %line (red) shows the maximum convergence radius $R$ as a function of density
   %$n$ while the lighter line (green) shows the optimal $\alpha$.  \textbf{(e)}
   %and \textbf{(f):} The pole maps of the diagMC series in the atomic limit in
   %complex $U$ plane.  Crosses indicate where we evaluate the series in the
   %corresponding Hubbard model ($\xi=1$).
\end{figure*}

We study the Hubbard model on an (infinite) square lattice: 
\begin{equation}
  \mathcal{H} = -\sum_{ i j  \sigma} t_{ij}\, c^\dagger_{i\sigma} c_{j\sigma}
    + U \sum_i n_{i\uparrow} n_{i\downarrow}, \label{Hubbard}
\end{equation}
%where $i,j$ label sites of an infinite square lattice, $c^\dagger_{i \sigma}$
%creates an electron on site $i$ with spin $\sigma$ and $n_{i\sigma} =
%c^\dagger_{i\sigma}c_{i\sigma}$. In the following, the only non-vanishing
%$t_{ij}$ are the nearest neighbour hopping $t$ and the next-nearest neighbour hopping $t'$. 
with hopping amplitudes $t$ and $t'$ between nearest-neighbour and next nearest-neighbour sites respectively, and use $t=1$ as our energy unit. 
In essence, DiagMC is an efficient way of computing the coefficients $a_l$ of a perturbative series for
the self-energy as a function of the Matsubara frequency $\omega_n$ and
momentum $\mathbf{k}$:
\begin{equation}
  \Sigma(i\omega_n, \mathbf{k}) = \lim_{L \to \infty} \sum_{n=1}^{L} a_l(i\omega_n, \mathbf{k}) U^l,
  \label{eq:series}
\end{equation} 
where $a_l$ is a sum of all one particle irreducible (1PI) Feynman diagrams with $l$ interaction vertices 
connected  by non-interacting Green's functions 
$G_0(i\omega_n, \mathbf{k}) = [i\omega_n - \epsilon_\mathbf{k} + \mu]^{-1}$~\cite{AGD}. 
The success of this approach fundamentally relies on the convergence properties of the series~(\ref{eq:series}).
Because $a_l$ are stochastically computed in DiagMC as the sums of factorial (in $l$) number of sign-alternating contributions, in practice one can only reach $L=6\sim7$ with reasonable statistical error bars due to the fermionic
sign problem. Thus, a controlled answer is
only warranted  as long as the series~(\ref{eq:series}) can be reliably
extrapolated to $L \to \infty$ given an essentially limited number of computed expansion orders. This
extrapolation becomes increasingly difficult at low $T$ and large $U$.

In order to establish control over the convergence properties of the series, we
introduce a modified action $S_\xi$ given by
\begin{eqnarray}
  &S_{\xi} =  -\sum_{\omega_{n},\mathbf{k},\sigma}    c_{\omega_{n},\mathbf{k},\sigma}^{\dagger}\tilde{G_{0}}(i\omega_{n},\mathbf{k})^{-1}c_{\omega_{n},\mathbf{k},\sigma} \\
\label{eq:action}
 & - \xi \sum_{\omega_{n},\mathbf{k},\sigma} \alpha_{\mathbf{k}}(i\omega_n)  c_{\omega_{n},\mathbf{k},\sigma}^{\dagger} c_{\omega_{n},\mathbf{k},\sigma}
  + \xi U\int_{0}^{\beta}n_{\tau\uparrow}n_{\tau\downarrow}d\tau,
\nonumber
\end{eqnarray}
where $\tilde{G}_0(i\omega_{n},\mathbf{k})^{-1} = i\omega_n + \mu - \epsilon_{\mathbf{k}} -
\alpha_{\mathbf{k}}(i\omega_n) = G_0(i\omega_n,\mathbf{k})^{-1} - \alpha_{\mathbf{k}}(i\omega_n)$ and
$\alpha_{\mathbf{k}}(i\omega_n)$ is an arbitrary auxiliary field (cf.~\cite{rubtsov2005,profumo2015,Rossi2016shifted_action});
at $\xi=1$ we recover the original action for the Hamiltonian~(\ref{Hubbard}).
Expanding %the action with regard to  the parameter 
in powers of $\xi$ and applying Wick's
theorem generates a new diagrammatic representation of $\Sigma$:
\begin{equation}
  \Sigma(i\omega_n, \mathbf{k}) = \lim_{M \to \infty} \sum_{m=1}^{M} \tilde{a}_{m}(i\omega_n, \mathbf{k}) \xi^m,
  \label{eq:series-modified}
\end{equation} 
Here the coefficient $\tilde{a}_m$ is a sum of all the diagrams for $a_l$ with $l$ running from $1$ to $m$, and for each diagram of order $l$, an additional sum over all possible ways of inserting $m-l$ instances of
$\alpha_{\mathbf{k}}(i\omega_n)$ in the fermionic lines is performed according to the
standard diagrammatic rules of expanding with respect to an external field~\cite{AGD}.
Note that here the bare propagators $G_0(i\omega_n, \mathbf{k})$ are now replaced with
$\tilde{G_0}(i\omega_n, \mathbf{k})$.
The freedom in choosing $\alpha_{\mathbf{k}}(i\omega_n)$ can be now used to control the
convergence properties of the modified series~(\ref{eq:series-modified}) at
$\xi=1$.   

This freedom comes at the expense of extending the diagrammatic space, which considerably worsens the sign problem. To make practical calculations feasible we introduce a recursive protocol for summing all the diagrams for $\tilde{a}_m$, which makes use of the already computed $\tilde{a}_n$, $n<m$. The idea is that a high-order diagram may contain (between propagator lines) self-energy insertions of lower orders; all possible insertions of the total order $p$ can be implicitly summed and integrated over the internal momentum/frequency variables by including the results for $\tilde{a}_n$, $n \leq p$ in the propagator lines:
\begin{equation}
  G^{(p)} = \sum_{n=0}^{p} G^{(n)} \tilde{a}_{p-n} \xi^{p-n} \tilde{G}_0,
  \;\;\; p>0, \;\;\; G^{(0)} \equiv \tilde{G}_0,
\label{G-dressed}
\end{equation}
so that $G^{(p)} \propto \xi^p$. Then $\tilde{a}_m$ can be obtained by DiagMC sampling of only
1PI skeleton diagrams of order $l=\{1, \ldots, m\}$, where in each diagram some bare propagators $\tilde{G_0}$ are randomly replaced by dressed propagators $G^{(p_i)}$ so that $\sum_i p_i = m-l$. 
This recursive approach substantially improves the efficiency of DiagMC by effectively reducing the configuration space and can be generalized to other channels, e.g. by introducing dressed interaction lines $W^{(p)}$, dressed
two-particle irreducible vertices $\Gamma^{(p)}$, etc.

\paragraph{Illustrative result at high $T/t$.} 

We first investigate the simplest case of a constant field $\alpha_{\mathbf{k}}(i\omega_n) \equiv\alpha$.
In Fig.~\ref{fig:convergence}, we illustrate its effect on the Hubbard model at $U=4$ and $T=0.5$, 
using determinantal QMC simulation on a $16 \times 16$ lattice as a benchmark~\cite{bss1981}.  
In the first row of Fig.~\ref{fig:convergence}, we compare the value of $\Sigma(\mathbf{k}, i\omega_n)$ at the first few Matsubara frequencies and $\mathbf{k}= (\pi/4,\pi)$ summed up to order $K$, i.e. $\Sigma(\mathbf{k}, i\omega_n) = \sum_{m=1}^K
\tilde{a}_m(\mathbf{k},n) \xi^m$. Fig.~\ref{fig:convergence}a shows the behaviour of the standard series (\ref{eq:series}) (with the Hartree diagrams included in the Green's function following Refs. \cite{proceedings,kozik2010diagmc}), 
Fig.~\ref{fig:convergence}b and Fig.~\ref{fig:convergence}c show the behavior
for two different choices of $\alpha$. Clearly, the standard series and the one for an arbitrarily selected $\alpha=0.6$ fail to converge within accessible orders.  However, a clever choice of $\alpha =1.53 $ yields a great
improvement of convergence. The exact result is
recovered already at order $4$ and the extrapolation of the series to infinite order is straightforward.

\paragraph{Rationale: pole-moving.}

In order to get insight into the improvement brought by the introduction of a
modified action, we study in details the limiting case $t=0$, the Hubbard
atom, which can be solved exactly. In particular, we show how tuning $\alpha$
allows to control the convergence radius of the series~(\ref{eq:series-modified}).
The self-energy for the action $S_\xi$ and $t=0$ is given by
\begin{align}  
  \Sigma(i\omega_n) &= \frac{n\xi U}{2} + \frac{1}{4}
  \frac{n(2-n)\xi^2 U^2}{i\omega_n + \tilde{\mu} - (2-n)\xi U/2} \\
  \tilde{\mu} &= \xi \alpha - \alpha + \mu
\end{align}
where $n = [e^{\beta\tilde{\mu}} +e^{2\beta \tilde{\mu}-\beta \xi U }]/[1+2e^{\beta
\tilde{\mu}}+e^{2\beta \tilde{\mu}-\beta\xi U }]$ is the density. The analytical
structure of $\Sigma(i\omega_0)$ in the complex-$\xi$ plane is shown in
Fig.~\ref{fig:convergence}e and Fig.~\ref{fig:convergence}f. The convergence
radius $R$ of the series expansion in $\xi$ is given by the distance from the
origin to the closest pole in the complex-$\xi$ plane, which strongly depends on
the value of $\alpha$. For $\alpha = 0.6$ a pole is closer to the origin
than the evaluation point $\xi = 1$ and the series diverges, whereas for
$\alpha = 1.53$ the poles are further away and the series is convergent at $\xi
=1$. When $\alpha$ is further increased, new poles get closer to the origin and
there is therefore an optimal value for $\alpha$ for which the radius of
convergence is maximal. 
A systematic study for the full Hubbard model at $T=0.5$ suggests an optimal value of 
$\alpha\simeq1.53$, close to this atomic estimate  $\alpha \sim 1.3$ , as expected from a similar 
analytic structure of $\Sigma$ at this high temperature. 
Thus the Hubbard atom can provide a reasonable guide for finding the optimal $\alpha$.  
Finally, we find the largest convergence radius and the
corresponding optimal $\alpha$ for different densities of the Hubbard atom, as displayed in 
Fig.~\ref{fig:convergence}d. We see that $R$ is infinite at half-filling and
becomes finite ($R \lesssim 2.5$) as soon as a doping is introduced. For $U=4$,
the convergence radius is always large enough for the series to converge. It
has a minimum $R \simeq 1.6 > 1$ around $10\%$ hole (or electron) doping.

\paragraph{Reaching the pseudogap scale.}

\begin{figure}[!t]
 \begin{center}
\includegraphics[width=0.96\columnwidth, height= 0.7\columnwidth ]{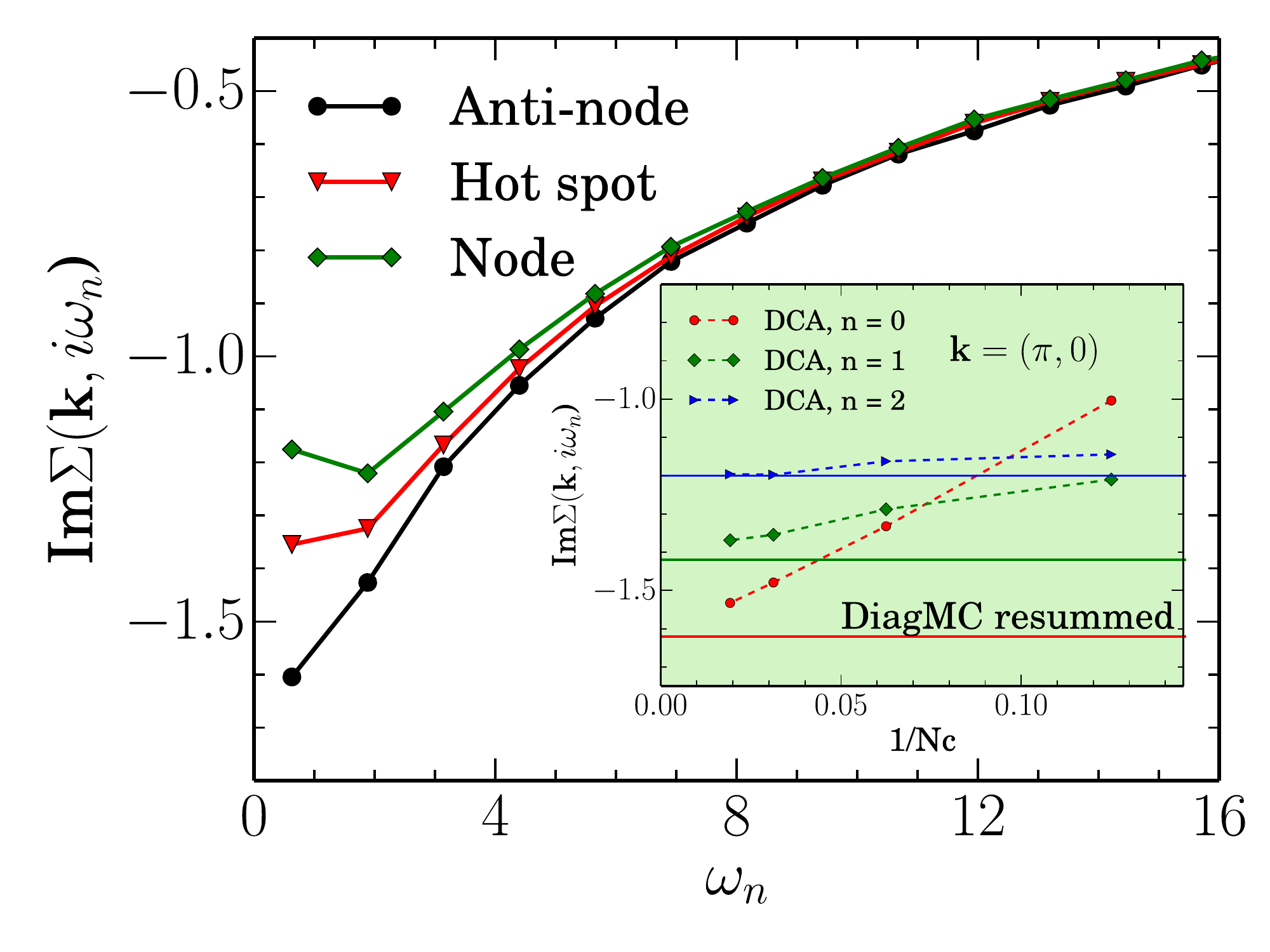}
   \end{center}
   \caption{\label{fig:imSigma}
     Imaginary part of the self-energy at the node, hot-spot  and anti-node
      at $U=5.6$, $t^{\prime}=-0.3$, $n=0.96$, $T=0.2$. Inset: DCA
     results with cluster size $N_{c}=8$, $16$, $32$, $52$ extrapolate to the
     DiagMC-summed result at different frequencies.}
     % as cluster size $N_c$ goes to infinity.
     % At small frequencies a prominent
     % nodal/anti-nodal differentiation is observed.
\end{figure}

We now show that this improved scheme allows one to reach the pseudogap
region~\cite{tremblay2006pseudogap,sordi2012, Macridin2006,gull2013}. We consider the Hubbard model
at $4\%$ hole doping and $U=5.6$, $t^{\prime} = -0.3$. 
We could achieve convergence down to $T=0.2$, where we compute the self-energy up to 7\textsuperscript{th} order with an optimized $\alpha = 2.3$. In
Fig.~\ref{fig:imSigma}, we display the imaginary part of the self-energy
$\mathrm{Im} \Sigma(\mathbf{k}, i\omega_n)$ taken at three different momenta $\mathbf{k}$ on the Fermi surface (FS). 
We see that the self-energy behaves differently at the nodal point $\mathbf{k}_N=(1.47,1.47)$ (intersection of the FS 
with the zone diagonal) in comparison to the antinode $\mathbf{k}_{AN}=(3.04,0.49)$ (where the FS hits the upper zone boundary). 
The imaginary part of the AN self-energy extrapolates to a larger negative value at low-frequency, indicating strongest correlation effects at the AN. 
%
%at the node $K_N=(1.47,1.47)$, at the antinode $K_{AN}=(3.04,0.49)$ and 
%also at the `hot-spot' corresponding to the intersection of the Fermi surface with the 
%antiferromagnetic zone boundary \ag{$K_{HS}= ??$}. We see that the two self-energies have a different
%behavior: the antinodal self-energy extrapolates to a larger negative value at
%small frequencies, indicating that the effects of correlations are stronger at the antinode than at the node. 
%
Hence, a clear N/AN differentiation is already apparent at $T=0.2$, consistently with previous calculations~\cite{Macridin2005,tremblay2006pseudogap} indicating 
that this temperature coincides with the onset of the pseudogap at $U=5.6$. 
%
%The inset of Fig.~\ref{fig:imSigma} compares our results (solid lines) with DCA calculations for different system sizes. 
%We observe that the extrapolation of the DCA is in excellent agreement with our results.
The inset of Fig.~\ref{fig:imSigma} also demonstrates that our results at the AN are in excellent agreement with 
large scale dynamical cluster approximation (DCA)\citep{maier2005dca} ones (after extrapolating the latter as a function of cluster size). 
Finally, we note (Fig.~\ref{fig:imSigma}) that the self-energy is larger at the AN than at the `hot-spot' $\mathbf{k}_{HS}=(2.26,0.88)$ 
(intersection of the FS with the antiferromagnetic zone boundary),  
indicating that we have reached a regime in which the weak-coupling spin-fluctuation picture does not apply. 
Being able to resolve the difference of behaviour at the HS and AN is a clear advantage of the current approach as compared 
to cluster methods. 

\paragraph{Physical origin of the nodal/antinodal dichotomy: 
antiferromagnetic spin correlations.}

\begin{figure}[!t]
 \begin{center}
\includegraphics[width=0.98\columnwidth, height= 0.5\columnwidth ]{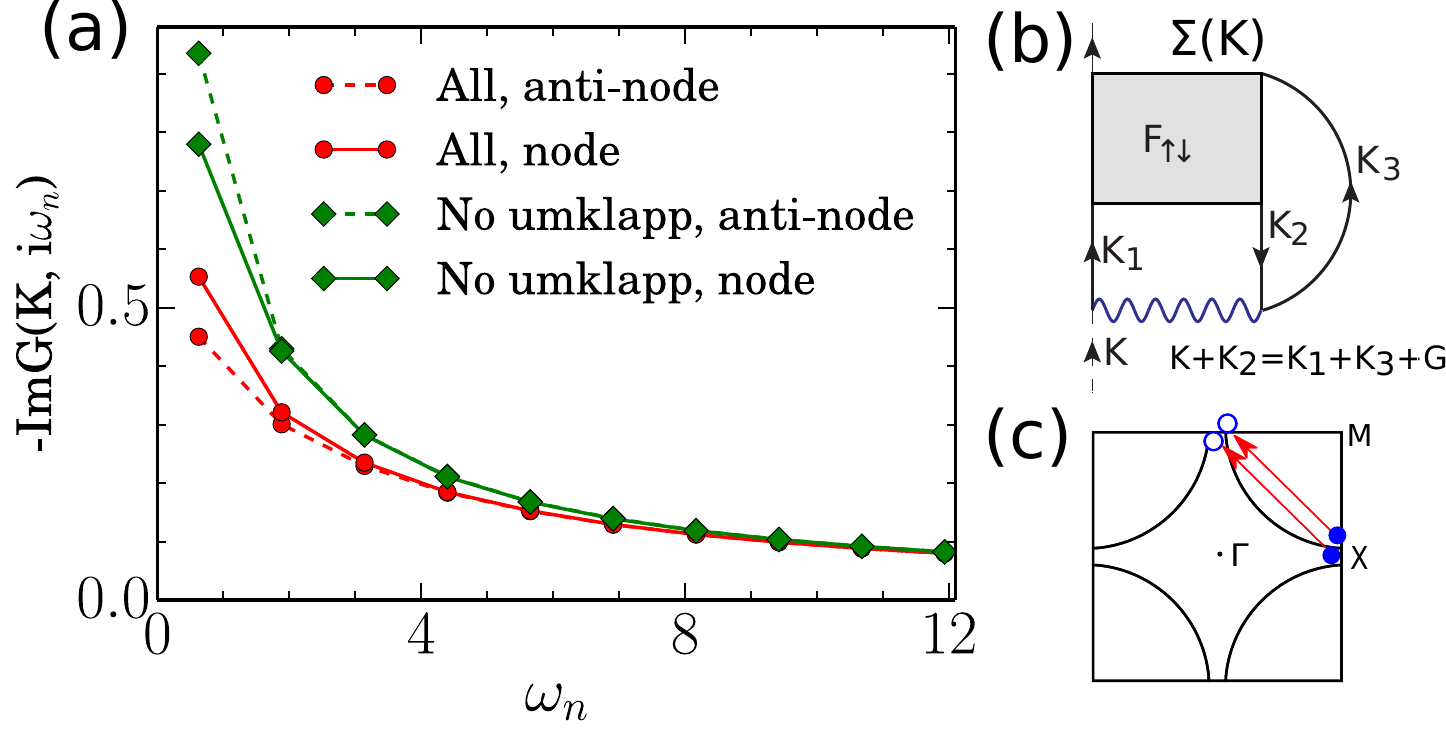}
   \end{center}
   \caption{\label{fig:imG}
     \textbf{(a)}: Imaginary part of the Green's functions at node and
     anti-node as a function of frequency,  in cases with and without umklapp
     processes.  \textbf{(b)}: Decomposition of the self-energy using the Dyson-Schwinger equation of
     motion, \textit{c.f. Eq.}~\ref{eq:EoM}. \textbf{(c)}:  Illustration of a typical weak-coupling umklapp process with
     $G=(2\pi,2\pi)$. }
     %The momentum conservation of the vertex is fulfilled up to a reciprocal
     %vector $G$.  The umklapp scattering process corresponds to $G\neq0$.
     %In the case with all scattering processes taken into account, the
     %anti-nodal Green's function is more suppressed than the nodal one. When
     %the umklapp  scattering is deliberately removed, the anti-nodal Green's
     %function becomes more metallic than the nodal Green's function.
\end{figure}
A decisive asset of DiagMC is that it provides direct information about
the mechanisms behind the pseudogap and N/AN differentiation.  
We demonstrate that umklapp processes are essential to the destruction of the AN quasiparticles. 
To this aim, we decompose the self-energy as shown in Fig.~\ref{fig:imG}b and monitor the momentum entering the two-particle scattering amplitute $F_{\uparrow \downarrow}$ during the DiagMC evaluation. By forcing the sum of incoming and outcoming momenta of $F_{\uparrow \downarrow}$ to differ by a non-zero or zero reciprocal lattice vector $G$, we allow or forbid umklapp scattering at will. 
%Momentum conservation requires the sum of incoming momenta and outcoming ones to be equal up to a reciprocal lattice vector $G$. 
%Umklapp processes correspond to $G \ne 0$. 
%where we display the imaginary part of the Green's function at the node and at the antinode with and without umklapp processes.  
The results are given in Fig.~\ref{fig:imG}: when umklapp processes are forbidden, both the imaginary part of N and AN 
Green's functions become significantly larger, indicating that umklapp processes are 
relevant to suppress spectral weight in the full solution. More
importantly, without umklapp processes AN Green's function turns out
to be more coherent than the N one, while the opposite is true when umklapp
processes are allowed.  They are therefore a key ingredient in the suppression
of spectral weight at the AN, and their importance is actually found to grow as the perturbation order increases (not shown).

\begin{figure*}[!t]
 \begin{center}
\includegraphics[width=0.8\textwidth, height=0.35\textwidth]{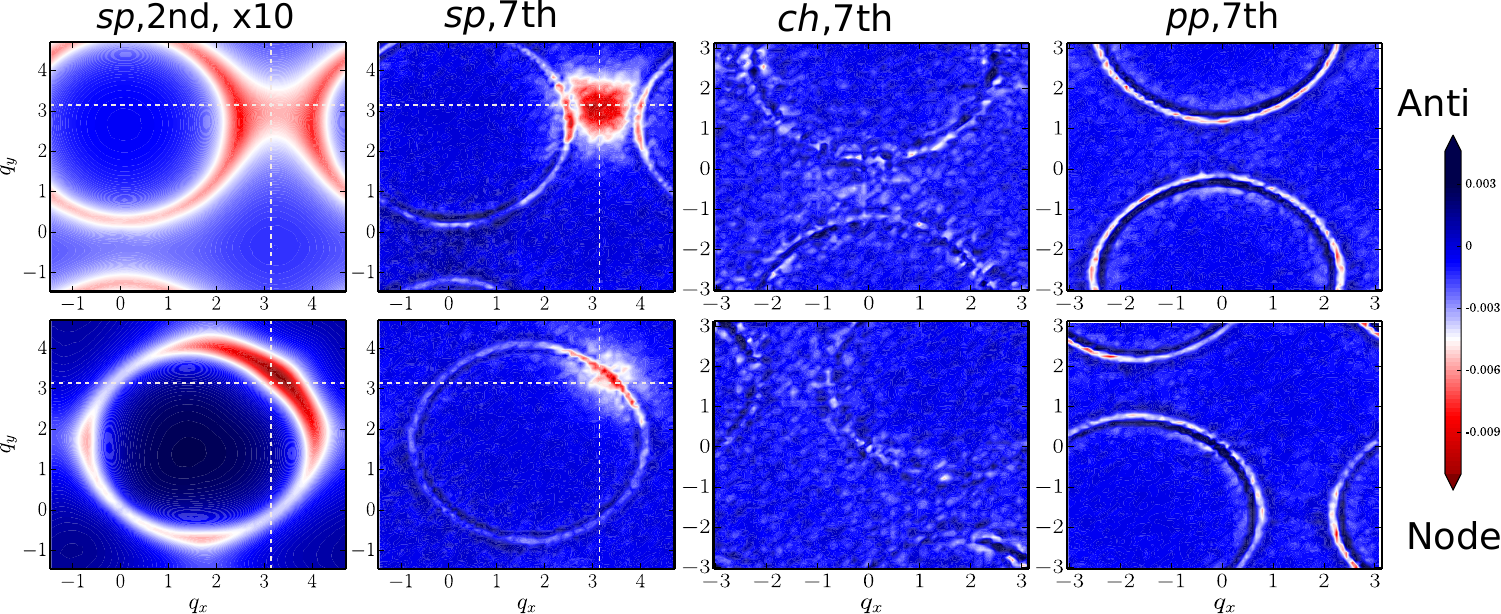}
   \end{center}
   \caption{\label{fig:fluctuation_diagnostics}
   Maps of the transfer momentum $\mathbf{q}$ contributions to
   $\mathrm{Im}\Sigma(\mathbf{k},i\omega_0)$ + $\mathrm{Im}\Sigma(\mathbf{k},i\omega_1)$ in the spin (sp), charge (ch) and
   particle-particle (pp) channels, both at the node, $\mathbf{k}=(1.47,1.47)$ (bottom row) and
   anti-node, $\mathbf{k}=(3.04,0.49)$ (top row). The two panels on the left are results
   at perturbation order 2 while the rest are obtained up to order 7.}
\end{figure*}

To analyse further the N/AN dichotomy, we employ the Dyson-Schwinger
equation representing the self-energy at a given momentum $\mathbf{k}$ as (Fig.~\ref{fig:imG}b): 
\begin{equation}
  \Sigma(\mathbf{k}) = \frac{Un}{2} - \sum_{\mathbf{k'},\mathbf{q}} F_{\uparrow \downarrow}(\mathbf{k},\mathbf{k'},\mathbf{q}) g(\mathbf{k}) g(\mathbf{k'}+\mathbf{q}) g(\mathbf{k}+\mathbf{q}),
  \label{eq:EoM}
\end{equation}
%For the Hubbard model we can decompose the self-energy as  
or $\Sigma(\mathbf{k}) = \frac{Un}{2} -\sum_{\mathbf{q}}[\Sigma^{\mathbf{q}}_{X}(\mathbf{k})]$, and study the contributions from different collective modes with the transfer momentum $\mathbf{q}$, 
where $X$ stands for $\textit{spin} $ (sp), $\textit{charge}$ (ch) or $ \textit{particle-particle}$ (pp) representations, 
 %\textit{ spin}, \textit{charge} and \textit{particle-particle}
\begin{eqnarray}
\Sigma_{sp}^{\mathbf{q}}(\mathbf{k})&=&\sum_{\mathbf{k'}} -F_{\uparrow\downarrow}(\mathbf{k},\mathbf{k}+\mathbf{q},\mathbf{k'}-\mathbf{k})g(\mathbf{k'})g(\mathbf{k'}+\mathbf{q})g(\mathbf{k}+\mathbf{q})\notag \\
\Sigma_{ch}^{q}(\mathbf{k})&=&\sum_{\mathbf{k'}} [F_{\uparrow\downarrow}(\mathbf{k},\mathbf{k}+\mathbf{q},\mathbf{k'}-\mathbf{k}) \\
& & -2\times F_{\uparrow\downarrow}(\mathbf{k},\mathbf{k'},\mathbf{q})]g(\mathbf{k'})g(\mathbf{k'}+\mathbf{q})g(\mathbf{k}+\mathbf{q}) \notag \\
\Sigma_{pp}^{\mathbf{q}}(\mathbf{k})&=&\sum_{\mathbf{k'}} -F_{\uparrow\downarrow}(\mathbf{k},\mathbf{k'},\mathbf{q}-\mathbf{k}-\mathbf{k'})g(\mathbf{k'})g(\mathbf{q}-\mathbf{k'})g(\mathbf{q}-\mathbf{k})\notag
\end{eqnarray}
The above equations are obtained by expressing $F_{sp/ch/pp}$ in terms of $F_{\uparrow \downarrow}$ (see \cite{gunnarsson2015fluct}).
The resulting low-energy intensity maps of $\Sigma^{\mathbf{q}}_{X}(\mathbf{k})$  (obtained at order 7) for all three representations at the N 
and at the AN are displayed in Fig.~\ref{fig:fluctuation_diagnostics}. 
The first compelling result is that the charge and the particle-particle representations are essentially featureless
for both the N and AN self-energies. The only visible patterns
stem from FS to FS transfer momenta, as shown by the light circles. 
%
%In contrast, scattering in the spin channel makes the largest contribution to the imaginary part of the self-energy 
%at low energies.
In contrast, the spin channel exhibits dominant modes responsible for 
the largest contribution to the imaginary part of the self-energy at low-energies.  
Interestingly, the most significant transfer momenta are all close to $(\pi, \pi)$ and we conclude that
antiferromagnetic spin correlations are the leading scattering mechanism in the
pseudogap region of the phase diagram, in agreement with recent experimental findings~\cite{badoux2016pseudo}.  
The fine momentum resolution of DiagMC allows us to examine the difference between
the N and AN self-energy in further detail.  We see that, at the
N the transfer momenta are close, but not exactly at
$(\pi,\pi)$. They are concentrated around the Fermi surface and are not exactly
commensurate. On the contrary, at the AN the transfer
momenta are centered around $(\pi,\pi)$ and the corresponding amplitude is larger. 
The existence of a saddle-point in the band dispersion (van-Hove singularity) 
close to the AN may be at the origin of this difference, the flatter dispersion allowing 
to compensate for the energetic cost of commensurate $(\pi,\pi)$ spin scattering. 
The relevance of the saddle-point has been discussed in Refs.\cite{lehur2009review,furukawa1998, honerkamp2001weakumklapp} 
within weak-coupling approaches.   
%
%We believe that the reason for this difference in behavior for the
%nodal and antinodal self-energy is due to the presence of the van Hove
%singularity close to the antinodal point. The energy landscape is flatter close
%to the antinode and the energetic price can be compensated to have fully
%commensurate $(\pi,\pi)$ spin scattering.  At the node instead the energy
%landscape is steeper and it is too costly energetically to allow commensurate
%scattering: the momentum transfer remains on the Fermi surface. 
%
Importantly, we note (Fig.~\ref{fig:fluctuation_diagnostics}) that FS scattering involving an incommensurate 
transfer momentum controls the nodal self-energy at both low and high perturbation orders. In contrast, 
$(\pi,\pi)$ scattering emerges at high perturbation orders at the AN. Hence, the scattering mechanism remains of the 
weak-coupling type at the N, while the pseudogap opening at the AN is a strong-coupling phenomenon for the 
value of $U/t$ studied here. 
%
%It's worthy noting another essential difference between the antinodal and nodal physics, which is revealed by a
%comparison of $\mathrm{Im}\Sigma^{q}_{\mathrm{sp}}(K)$ at second and
%7\textsuperscript{th} order, see Fig.~\ref{fig:fluctuation_diagnostics}.  At
%the node, the transfer momenta contributing to the self-energy connect Fermi
%surface points both at low and high orders. The mechanism at the node remains
%of the weak-coupling kind ~\cite{furukawa1998umklapp}. On the contrary, at the
%antinode the dominant transfer momenta near $(\pi,\pi)$ at
%7\textsuperscript{th} order are absent at second order. This clearly shows that
%the nodal/antinodal dichotomy and the opening of the pseudogap are strong-coupling effects.

\paragraph{Conclusions and perspectives.} 

In this letter, we have introduced an improved DiagMC method relying on an optimized parametric modification of the Hubbard model action. 
This allows us to access in a controlled way and with high momentum resolution parameter regimes that were previously unreachable, such as the onset of the pseudogap and nodal/antinodal differentiation.  
We show that these effects are due to antiferromagnetic correlations and that marked differences with weak-coupling 
spin-fluctuation theories appear in the regime of coupling investigated here. The challenge ahead is to improve the current method 
in order to reach significantly lower temperatures. 

\begin{acknowledgments}
We would like to thank P. J. Hirschfeld , A. J. Millis, O. Parcollet, N. Prokof'ev, B. Svistunov, Ning-Hua Tong and 
A.-M.S. Tremblay, for useful discussions. 
This work has been supported by the Simons Foundation within the Many Electron Collaboration framework. 
We also acknowledge support of the European Research Council (ERC-319286 QMAC) and of the 
Swiss National Supercomputing Center (CSCS) under project s575. Some of the calculations were
performed with the TRIQS~\cite{triqs2015} toolbox.
\end{acknowledgments}

\bibliography{diag}

\end{document}